\documentclass[%
 reprint,
 amsmath,amssymb,
 aps,
]{revtex4-2}

\usepackage{graphicx}% Include figure files
\usepackage{dcolumn}% Align table columns on decimal point
\usepackage{bm}% bold math
\newcommand{\updatedfigure}[3]{%
\IfFileExists{#1}{\includegraphics[width=\linewidth]{#1}}{%
\fbox{\parbox[c][#2][c]{\dimexpr\linewidth-2\fboxsep-2\fboxrule\relax}{%
\centering\small #3\par\medskip\texttt{\detokenize{#1}}}}}}

\begin{document}

\preprint{APS/123-QED}

\title{Semi-Analytical Geometric Synthesis of Robust Quantum Controls}

\author{Eduardo Willwock Lussi}
 \altaffiliation[Also at ]{Quantuloop, Salvador, Brazil.}
\author{Rafael de Santiago}
\affiliation{Department of Informatics and Statistics, Federal University of Santa Catarina, Florianópolis, Brazil.}

\author{Eduardo Inacio Duzzioni}
\affiliation{Department of Physics, Federal University of Santa Catarina, Florianópolis, Brazil.}

\author{Hector Bessa Silveira}
\affiliation{Department of Automation and Systems Engineering, Federal University of Santa Catarina, Florianópolis, Brazil.}

\date{\today}

\begin{abstract}
High-fidelity quantum gates are a central requirement for fault-tolerant quantum computation, yet their implementation is limited by the sensitivity of qubits to environmental noise and by control non-idealities. To mitigate these effects, quantum gates can be designed as dynamically corrected gates (DCGs), where control sequences are engineered to suppress the impact of errors on the resulting operation. Many methods in the literature are able to synthesize DCGs, but they typically rely on numerical optimization, with synthesis times in the order of minutes per gate. Here, we propose an automated semi-analytical method that reduces this cost by orders of magnitude, synthesizing robust smooth control pulses in the order of tens of milliseconds rather than minutes, with comparable fidelity and robustness. The present work relies on the Space Curve Quantum Control (SCQC) formalism, where quantum evolution on $\mbox{SU}(2)$ is mapped on a space curve whose geometric properties determine the target gate and its robustness. Based on the SCQC design framework, the new approach proposed in this paper consists in formulating the robust gate generation problem for single-qubit gates as an optimal control problem in which the necessary optimality conditions established by the classical Pontryagin's Maximum Principle (PMP) determine a two-point boundary value problem whose numerical solution allows one to efficiently construct smooth controls that are robust against both amplitude and dephasing errors. The corresponding optimal state and costate equations of PMP are analytically integrated using Rodrigues' rotation formula, and solved using a shooting method. Starting from a reference solution, we gradually vary the target rotation angle to construct a library of solutions, whose interpolation provides initial guesses for the shooting parameters of new target gates. We evaluated the method on 2,000 single-qubit rotations, obtaining a mean local optimization time of $39.945$~ms per gate after preparation of the interpolation base. Quantum propagation yielded nominal gate infidelities under $10^{-9}$ after virtual phase correction, with numerically verified robustness against quasistatic amplitude and dephasing errors.
\end{abstract}

\maketitle

\section{\label{sec:introduction}Introduction}

Quantum error correction is a fundamental ingredient for scalable quantum computation, providing a means to protect logical quantum information from errors affecting physical qubits \cite{Gottesman2009}. Fault-tolerant quantum computation requires physical error rates to remain below a code- and noise-dependent threshold, such that increasing the size of the error-correcting code can progressively suppress logical errors \cite{Hashim2023, Bravyi2024, Acharya2025}. Since quantum gates are the fundamental operations used to manipulate quantum information, achieving sufficiently low gate error rates is therefore essential for realizing fault-tolerant quantum computation.

Quantum errors can arise from both unwanted interactions between the qubit and its environment and non-idealities in the applied control fields \cite{Tripathi2025, Hashim2023}. Environmental noise can induce energy relaxation and dephasing, limiting the coherence and lifetime of a qubit. At the same time, non-idealities in the control fields, such as errors in amplitude, frequency, phase, or pulse duration, can cause the implemented operation to deviate from the desired gate. Additional errors, such as leakage out of the computational subspace and crosstalk between qubits, can further degrade the fidelity of quantum operations.

Quantum control can be used to design control fields that mitigate the effects of noise and control non-idealities while implementing a desired quantum operation \cite{Koch2022}. Different control strategies can be used to address this problem. Analytical pulse-shaping techniques, such as DRAG, can be designed to mitigate errors associated with weak anharmonicity, including leakage and phase errors \cite{Motzoi2009, Gambeta2011}. Composite pulses \cite{Wimperis1994} employ sequences of control pulses to compensate for specific systematic errors. More general approaches, such as optimal and robust control, provide adequate frameworks for designing control protocols according to prescribed performance objectives. Optimal control seeks control protocols that optimize a given objective, such as gate fidelity, operation time, or control energy \cite{Koch2022}, whereas robust control aims to maintain the desired performance in the presence of parameter uncertainties and perturbations \cite{Weidner2025}.

Among robust control strategies, dynamically corrected gates (DCGs) have emerged as a means of implementing robust quantum operations. Unlike conventional dynamical decoupling \cite{Viola1998, Viola1999}, whose primary goal is to preserve quantum states by suppressing unwanted system-environment interactions, DCGs are designed to perform a target gate while simultaneously canceling the leading-order contributions of specified errors \cite{Khodjasteh2009a, Khodjasteh2009b}.

Several strategies have been proposed to construct DCGs under realistic experimental constraints. Analytical approaches range from dynamical-decoupling-inspired constructions with arbitrarily accurate error suppression under bounded-strength controls \cite{Khodjasteh2010} and filter-function-based pulse design \cite{Green2013}, to semi-analytical smooth-pulse parametrizations \cite{Gungordu2019} and inverse-optimization methods that construct robust gates from prescribed geometric trajectories \cite{Dridi2020}. In parallel, numerical optimal control methods -- including gradient-based algorithms \cite{Khaneja2005}, the chopped random basis technique \cite{Caneva2011}, and automated DCG synthesis under bounded-control constraints \cite{Khodjasteh2012} -- have since been extended into general frameworks for robustness against arbitrary, uncharacterized noise \cite{Poggi2024}, and validated experimentally on cloud-accessible \cite{Carvalho2021} and integrated superconducting hardware \cite{Yi2024}.

More recently, space curve quantum control (SCQC) has emerged as a promising framework for robust quantum control, providing a geometric perspective that recasts the design of DCGs as the construction of a space curve whose geometric properties encode the desired noise-robustness properties. This formalism was first introduced for single-qubit gates robust to quasistatic dephasing noise \cite{Zeng2019}, subsequently extended to multiqubit systems \cite{Buterakos2021}, and later generalized to enforce simultaneous robustness against multiple, independent noise sources \cite{Nelson2023}. A conceptually related geometric correspondence between noisy quantum dynamics and universally robust gate design has also been independently developed in Ref.~\cite{Hai2025}.

Among the methods built on SCQC, one closely related to our work is the Bézier Ansatz for Robust Quantum (BARQ) control method proposed in Ref.~\cite{Piliouras2026}. BARQ parametrizes the space curve using Bézier control points and numerically optimizes a subset of them to control amplitude errors and pulse smoothness, while the remaining control points are chosen to fix the target gate and enforce robustness against dephasing through boundary conditions on the curve in the adjoint representation. This decoupling of gate-fixing from noise-robustness eliminates the tradeoff between the two objectives that affects earlier SCQC-based constructions and yields smooth control pulses suitable for experiments. However, BARQ still relies on an iterative numerical optimization over the control points for each new target gate.

Efficient pulse synthesis is particularly important for dynamic gate generation, where controls are produced at compilation time according to each operation's execution context. In a previous work, we developed a time-optimization framework that exploits this opportunity by assigning fast implementations to operations on the circuit's critical path and exploiting scheduling slack to accommodate longer implementations without increasing the overall circuit latency \cite{Lussi2025}. Extending this strategy to the on-demand generation of robust pulses requires computationally efficient synthesis: repeatedly optimizing controls for individual gate instances can otherwise introduce substantial classical compilation overhead. Fast robust-pulse synthesis is therefore an important step toward generating controls that exploit the timing flexibility of each circuit, instead of relying exclusively on predefined waveforms.

 In this paper we propose a new technique for the robust gate generation problem for single-qubit gates. By relying on the SCQC design framework, the proposed approach consists in formulating the robust gate generation problem as an optimal control problem in which the classical Pontryagin's Maximum Principle (PMP) can be readily applied to determine necessary optimality conditions that result in a two-point boundary value problem. The corresponding optimal state and costate dynamics are analytically integrated using Rodrigues' rotation formula, so that the referred two-point boundary value problem is efficiently solved numerically using a shooting method \cite{StoerBulirsch2002}, thus constructing smooth controls that are robust with respect to both amplitude and dephasing errors. 

Closest in spirit to our approach is the method of Hanson and Lucarelli \cite{Hanson2024}, who employed PMP and shooting directly to $\mbox{SU}(n)$ evolution. Our SCQC-based formulation instead describes the space curve through its Frenet--Serret equations, taking curvature and torsion as state variables and their time derivatives as the controls. This choice yields a closed-form costate dynamics for a single-qubit and allows to analytically advance the frame by means of Rodrigues' rotation formula, while holding curvature and torsion constant within each integration interval.

Starting from a reference solution, we gradually vary the target rotation angle and follow the corresponding solutions using a tangent predictor, local shooting corrections, and adaptive angular steps \cite{BaderKunkel1989}. The resulting solutions define an interpolation of the shooting parameters, which provides initial guesses for local shooting corrections at new target angles. In a test comprising 2,000 single-qubit rotations, all queries satisfied the prescribed terminal and robustness tolerances, with a mean local optimization time of $39.945$~ms per gate after preparation of the interpolation base. Quantum propagation yielded nominal gate infidelities under $10^{-9}$. For comparison, the BARQ implementation available in the software package \texttt{qurveros} took on the order of minutes per gate on our computational platform (Intel Core i7-13620H, 32~GB of RAM, Windows).

The paper is organized as follows. Section~\ref{sec:scqc} describes the main ideas of SCQC-based design and their application to robust gate generation for a single-qubit. Section~\ref{sec:pmp} formulates the referred optimal control problem and derives the associated two-point boundary value problem, which in turn corresponds to the necessary optimality conditions determined by PMP. Section~\ref{sec:shooting} presents the shooting formulation and angular continuation strategy. Section~\ref{sec:validation} shows the numerical validation of the proposed framework, and Section~\ref{sec:conclusion} presents the conclusions.

\section{\label{sec:scqc}Space Curve Quantum Control}

The Space Curve Quantum Control (SCQC) formalism provides a geometric representation of single-qubit dynamics in which robustness to control non-idealities can be expressed as geometric constraints on a space curve \cite{Zeng2019}. The formalism was originally introduced for gates robust against quasistatic dephasing and was later extended to incorporate additional error mechanisms, including multiplicative control-amplitude errors \cite{Buterakos2021,Nelson2023}. In the sequel we summarize the main ideas of SCQC-based design, and refer the reader to Refs.~\cite{Zeng2019,Buterakos2021,Nelson2023} for further details. 

Consider the controlled single-qubit Hamiltonian
\begin{equation}\label{controlledhamiltonian}
H(t)
=
\frac{\Omega(t)(1 + \epsilon)}{2}
\left[
\cos\Phi(t)\sigma_x
+
\sin\Phi(t)\sigma_y
\right]
+
\frac{(\Delta(t) + \delta_z)}{2} \sigma_z,
\end{equation}
where $\Omega(t)$, $\Phi(t)$, and $\Delta(t)$ denote the driving amplitude, phase, and detuning, respectively, $\sigma_x, \sigma_y, \sigma_z$ are the usual Pauli matrices, and $\epsilon, \delta_z$ are quasistatic parameters corresponding to multiplicative amplitude and additive detuning errors, respectively.

The robust gate generation control problem is defined as follows: given any target gate $U_{\rm target}\in \mbox{SU}(2)$ and a prescribed time duration $T_g>0$, determine smooth controls $\Omega(t)$ and $\Phi(t)$, for $t \in [0,T_g]$, which ensure that the corresponding evolution $U(t) \in \mbox{SU}(2)$ of \eqref{controlledhamiltonian} with $U(0) = I$ is such that $U(T_g) \cong U_{\rm target}$ (with respect to average gate fidelity), up to a global phase as well as an ideal virtual $Z$ correction at $T_g$, for all sufficiently small $\epsilon, \delta_z$ (uniformly). For a single-qubit, the average gate infidelity is given by
\begin{equation}
\mathcal I=1-\frac{2+|\operatorname{Tr}(U_{\rm target}^\dagger U_{\rm corrected})|^2}{6}.
\label{eq:gate-infidelity}
\end{equation}
Here, $U_{\rm corrected} = R_Z(\alpha) U(T_g)$, where $R_Z(\alpha)$ is a virtual Z rotation about some adequate angle $\alpha$. The average gate fidelity is $1-\mathcal I$.

Now, consider the nominal Hamiltonian $H_0(t)$ associated to $H(t)$, that is, taking $\epsilon = \delta_z = 0$ in \eqref{controlledhamiltonian}. Fix $T_g > 0$, and let $U_0(t)\in \mbox{SU}(2)$, $t \in [0, T_g]$, denote the evolution of $H_0(t)$ with $U_0(0)=I$ for chosen controls $\Omega(t), \Phi(t)$. In the interaction picture defined by this evolution, the error Hamiltonian associated with the detuning error is given by
\begin{equation}\label{firstorderexp}
\widetilde{H}_{\delta_z}(t)
=
\frac{\delta_z}{2}
U_0^\dagger(t)\sigma_z U_0(t).
\end{equation}
The effect of this error can be characterized by the Magnus expansion of the interaction-picture evolution. Factoring out the quasistatic error parameter, the first-order Magnus term is given by
\begin{equation}\label{firstorderterm}
A_1(t)
\triangleq
\int_0^t
U_0^\dagger(t')\sigma_z U_0(t')\,dt'
=
\mathbf{r}(t)\cdot\boldsymbol{\sigma},
\end{equation}
which determines the associated space curve $\mathbf{r}(t)\in\mathbb R^3$ (row vector), for $t \in [0, T_g]$, where $\boldsymbol{\sigma}=(\sigma_x,\sigma_y,\sigma_z)$. By taking the derivative on both sides, we have that 
\begin{equation}\label{evolution-tangent}
U_0^\dagger(t)\sigma_z U_0(t)
=
\mathbf{T}(t)\cdot\boldsymbol{\sigma},
\end{equation}
where $\mathbf{T}(t)=\dot{\mathbf{r}}(t)$ is the tangent vector of $\mathbf r(t)$. Since $U_0(t)^\dagger\sigma_zU_0(t)$ is a Pauli operator with unit Bloch-vector norm, then $\mathbf{T}(t)$ is a unit vector, and hence the space curve 
$\mathbf r(t)$ is already parametrizes by arc-length.

This construction provides a direct geometric interpretation of first-order noise cancellation. For the quasistatic detuning error $\delta_z$, the first-order error vanishes at $t = T_g$ when
\begin{equation*}
\int_0^{T_g}\mathbf{T}(t)\,dt=0,
\end{equation*}
which is equivalent to requiring that $\mathbf r(t)$ is a closed-curve, that is, 
$\mathbf{r}(T_g)=\mathbf{r}(0)$. Thus, first-order robustness against detuning or dephasing  corresponds to a closed space curve $\mathbf r(t)$.

The same geometric construction can be extended to the multiplicative amplitude error $\epsilon$. In this case, the first-order cancellation condition at $t = T_g$ requires that
\begin{equation*}
\int_0^{T_g}
\mathbf{T}(t)\times\dot{\mathbf{T}}(t)\,dt
=
0.
\end{equation*}
Geometrically, this condition corresponds to a vanishing net oriented area of the curve $\mathbf r(t)$ traced by its tangent vector $\mathbf{T}(t)$. 

Therefore, simultaneous first-order robustness against detuning and amplitude errors is obtained by imposing the following two conditions:
\begin{equation}\label{robustnessconditions}
\int_0^{T_g}\mathbf{T}(t)\,dt=0,
\qquad
\int_0^{T_g}
\mathbf{T}(t)\times\dot{\mathbf{T}}(t)\,dt=0.
\end{equation}
This ensures that $U_0(T_g)$ is close to $U(T_g)$ for small $\epsilon, \delta_z$.

The geometry of $\mathbf r(t)$ also determines how the controls $\Omega(t), \Phi(t)$ evolve. Let $\mathbf N(t)$ and $\mathbf B(t)$ be the unit normal and binormal of $\mathbf r(t)$, and let $\kappa(t)$ and $\tau(t)$ denote its signed curvature and torsion, respectively. The dynamics of the Frenet--Serret frame $F(t) = \left[ \mathbf T(t)' \;\; \mathbf N(t)' \;\; \mathbf B(t)' \right]' \in \mbox{SO}(3)$ is given by the Frenet--Serret equations
\begin{equation}
\begin{aligned}
\dot{\mathbf T}&=\kappa\mathbf N,\\
\dot{\mathbf N}&=-\kappa\mathbf T+\tau\mathbf B,\\
\dot{\mathbf B}&=-\tau\mathbf N.
\end{aligned}
\label{eq:frenet-serret}
\end{equation}
By comparing the geometric evolution of the curve $\mathbf r(t)$ with the nominal Hamiltonian $H_0(t)$, it can be shown that
\begin{equation}\label{geometriccontrols}
\Omega(t)=\kappa(t),\qquad
\dot{\Phi}(t)-\Delta(t)=\tau(t).
\end{equation}
Thus, the curvature of $\mathbf r(t)$ sets the driving amplitude, while the torsion sets the phase rate relative to detuning. In particular, in the absence of a detuning control, i.e.\ $\Delta(t)=0$, the phase is obtained by integrating the torsion:
\begin{equation}\label{phasecontrol}
\Phi(t)=\Phi(0)+\int_0^t\tau(t')\,dt'.
\end{equation}
Hence, the controls can be seen as being  parametrized by the geometry of the space curve $\mathbf r(t)$.

It is immediate from \eqref{evolution-tangent} that we necessarily have that
\begin{equation}\label{initialtangent}
    \mathbf T(0) = (0, 0, 1),
\end{equation}
since $U_0(0) = I$, and $\mathbf T_g \cdot\boldsymbol{\sigma}=U_0^\dagger (T_g)\sigma_z U_0(T_g)$, where $\mathbf T_g = \mathbf T(T_g)$ is the terminal tangent vector at $t=T_g$. Let $S^2$ be the unit sphere in $\mathbb{R}^3$. The map $\mbox{U}(2) \ni V \mapsto \mathbf{S} \in S^2$ defined by $\mathbf S \cdot\boldsymbol{\sigma} = V ^\dagger \sigma_z V$ is not injective. Indeed, 
%any rotation about the $z$-axis (of the Bloch sphere) leaves $\mathbf S$ unchanged because it commutes with $\sigma_z$. However, 
any two $V_1, V_2 \in \mbox{U}(2)$ yield the same $\mathbf S \in S^2$ if and only if they differ by a $Z$ rotation (acting by left multiplication), up to a global phase. In particular, if the terminal tangent vector $\mathbf T_g$ is such 
\begin{equation}
    \mathbf T_g \cdot\boldsymbol{\sigma}
=U_{\rm target}^\dagger\sigma_z U_{\rm target},
\label{eq:target-tangent}
\end{equation}
then $U_0(T_g)$ and $U_{\rm target}$ differ by some $Z$ rotation, up to a global phase. Recall that any $V \in \mbox{U}(2)$ can be identified with the rotation matrix $R_V \in \mbox{SO}(3)$ defined element-wise by
\begin{equation*}
(R_V)_{ij}
=
\frac{1}{2}
\mathrm{Tr}\left(V^\dagger\sigma_i V \sigma_j\right).
\end{equation*}
It can be shown that 
\begin{equation}\label{correctionrotation}
R_Z(\theta-\Phi(T_g))R_{U_0}(T_g)
= R_{U_{\rm target}},
\end{equation}
where $R_Z(\theta-\Phi(T_g))$ is the $Z$ rotation about the angle $\theta - \Phi(T_g)$, with the angle $\theta$ is determined by
\begin{equation}\label{thetarotation}
 R_Z(\theta) = R_{U_{\rm target}} \mathcal{R}_F(0) \mathcal{R}_F(T_g)', 
\end{equation}
and $\mathcal{R}_F(t) \triangleq \left[-\mathbf B(t)' \;\; \mathbf N(t)' \;\; \mathbf T(t)' \right]'$. 
Therefore, as seen from \eqref{correctionrotation}, no constraint on $\Phi(T_g)$ is required, and it suffices to perform an adequate virtual $Z$ rotation on $U_0(T_g)$ in order to generate $U_{\rm target}$.

Now, assume that $\mathbf r(t)$, $t \in [0, T_g]$, is any given space curve with unit tangent vector $\mathbf T(t)$ satisfying \eqref{initialtangent}. Choose the controls $\Omega(t)$ and $\Phi(t)$ as in \eqref{geometriccontrols} and \eqref{phasecontrol}, respectively, for the controlled Hamiltonian \eqref{controlledhamiltonian} with $U(0) = I$ and assuming that $\Delta = 0$. Then, it can be shown that \eqref{firstorderterm} is met for such $\mathbf r(t)$. We thus conclude that whenever such space curve $\mathbf r(t)$ is constructed in a manner that \eqref{robustnessconditions} and \eqref{eq:target-tangent} are met, then the referred robust gate generation control problem is solved. 

\section{\label{sec:pmp}Optimal Control Problem Framework for SCQC-Based Design}

In this section we show how the SCQC approach described above can be formulated as an optimal control problem that characterizes any space curve $\mathbf r(t)$, $t \in [0, T_g]$ with unit tangent vector $\mathbf T(t)$ that satisfies \eqref{initialtangent}, \eqref{robustnessconditions} and \eqref{eq:target-tangent}. Afterwards, we show how Pontryagin's Maximum Principle along with a shooting method can be applied in order to numerically solve such optimal control problem efficiently. Consequently, a particular space curve $\mathbf r(t)$, $t \in [0, T_g]$ with the required geometric properties will have been constructed, thus solving the robust gate generation control problem here treated by choosing the controls $\Omega(t)$ and $\Phi(t)$ in \eqref{controlledhamiltonian} with $\Delta = 0$ as in \eqref{geometriccontrols} and \eqref{phasecontrol}, respectively.

\subsection{Optimal Control Problem Formulation}

First of all, fix $T_g > 0$ and $U_{\rm target} \in \mbox{SU}(2)$, and note that the Frenet--Serret equations \eqref{eq:frenet-serret} can be rewritten in matrix form as
\begin{equation}\label{f-smatrix}
\dot F(t)=A(\kappa(t),\tau(t))F(t),
\end{equation}
where
\begin{equation}\label{framematrix}
A(\kappa(t),\tau(t))=
\begin{pmatrix}
0 & \kappa(t) & 0\\
-\kappa(t) & 0 & \tau(t) \\
0 & -\tau(t) & 0
\end{pmatrix}.
\end{equation}
The key ingredient is to consider \eqref{f-smatrix} as a controlled state equation with state $F(t) \in \mbox{SO}(3)$ and controls $\kappa(t), \tau(t)$. However, in order to be able to constrain the endpoint values of $\kappa(t), \tau(t)$ at $t = 0$ and $t = T_g$ in the optimal control problem to be formulated (this will be justified later on), consider the extended state $\big(F(t), \kappa(t), \tau(t)\big)$ with 
\begin{equation}
\dot{\kappa}(t)=u(t),
\qquad
\dot{\tau}(t)=v(t),
\end{equation}
where $u(t), v(t)$ are the new controls.

Now, in order incorporate the robustness conditions \eqref{robustnessconditions}, we define the following auxiliary state variables
\begin{equation}
\mathbf{P}(t)=\int_0^t\mathbf{T}(t')\,dt'
\label{eq:P}
\end{equation}
and
\begin{equation}
\mathbf{Q}(t) \triangleq \int_0^t\mathbf{T}(t')\times\dot{\mathbf{T}}(t')\,dt' = 
\int_0^t\kappa(t')\mathbf{B}(t')\,dt',
\label{eq:Q}
\end{equation}
which by construction have as state equations
\begin{equation}
\dot{\mathbf{P}}(t) =\mathbf{T}(t),
\qquad
\dot{\mathbf{Q}}(t) = \kappa(t) \mathbf{B}(t).
\end{equation}
Therefore, the complete state vector is $\mathbf x(t) = \big(F(t), \kappa(t), \tau(t), \mathbf{P}(t), \mathbf{Q}(t)\big)$ and $\mathbf{u}(t) = \big(u(t), v(t)\big)$ is the control vector. By definition, the auxiliary states satisfy
\begin{equation}
\mathbf{P}(0)=\mathbf{Q}(0)=\mathbf{0}.
\end{equation}
And note that the robustness conditions in \eqref{robustnessconditions} are equivalent to the following terminal constraints at $t=T_g$:
\begin{equation}
\mathbf{P}(T_g)=\mathbf{0},
\qquad
\mathbf{Q}(T_g)=\mathbf{0}.
\end{equation}

From \eqref{initialtangent}, we have that $\mathbf T(0)=(0,0,1)$, and take $\mathbf T(T_g) = \mathbf{T}_g$ according to \eqref{eq:target-tangent}. We also impose that
\begin{equation*}
\kappa(0)=\tau(0)=\kappa(T_g)=\tau(T_g)=0.
\end{equation*}
This curvature condition ensures that the pulse amplitude $\Phi(t) = \kappa(t)$ also vanishes at both endpoints (recall \eqref{geometriccontrols}), thus avoiding a jump discontinuity when the pulse is extended to a null pulse, for instance. This is crucial in order to avoid high-frequencies in the spectrum of the pulse. And, when $\Delta=0$ in \eqref{controlledhamiltonian}, then the torsion condition above forces the phase rate $\dot{\Phi}(t) = \tau(t)$ to vanish at the endpoints (a numerically favorable choice, as discussed below).

Lastly, we define an objective functional in order to design smooth controls $u(t)=\dot{\kappa}(t), v(t)=\dot{\tau}(t)$ that result in a curvature $\kappa(t)$ and torsion $\tau(t)$ that are of relatively small amplitudes and slow-variying on the time interval $[0,T_g]$.
%while satisfying the target-gate and robustness conditions. 
Therefore, we seek to determine an optimal control $\mathbf u^*(t)=\big (u^*(t), v^*(t) \big)$, $t \in [0, T_g)$, that minimizes the following objective functional:
\begin{equation}\label{functional}
J=\displaystyle
\frac{1}{2}
\int_0^{T_g}
\Big[
w_\kappa\kappa^2(t)
+w_\tau\tau^2(t)
+w_u u^2(t)
+w_v v^2(t)
\Big] dt,
\end{equation}
where $w_\kappa$, $w_\tau \geq 0$ and $w_u, w_v > 0$. The weights $w_\kappa$ and $w_\tau$ penalize the magnitudes of the pulse amplitude $\Omega(t) = \kappa(t)$ and of the phase rate $\dot{\Phi}(t) = \tau(t)$, respectively, while $w_u, w_v$ penalize their variations in time.

In sum, the proposed optimal control problem for SCQC-based design is formulated as:
\begin{equation}
\begin{aligned}
\mathbf u^* &\in \operatorname*{arg\,min}_{\mathbf u} J,\\
\text{subject to}\quad
\dot F &= A(\kappa,\tau)F,\\
\dot\kappa &= u, \qquad \dot\tau = v,\\
\dot{\mathbf P} &= \mathbf T,\qquad
\dot{\mathbf Q} = \kappa\mathbf B,\\[2pt]
\mathbf T(0) &= (0,0,1),\\
\mathbf T(T_g) &= \mathbf T_g \quad \mbox{as in \eqref{eq:target-tangent}},\\
F(0),\,F(T_g) &\in \mathrm{SO}(3),\\
\kappa(0) &= \tau(0) = 0,\\
\kappa(T_g) &= \tau(T_g) = 0,\\
\mathbf P(0) &= \mathbf Q(0) = \mathbf 0,\\
\mathbf P(T_g) &= \mathbf Q(T_g) = \mathbf 0.
\end{aligned}
\label{eq:scqc-optimal-control}
\end{equation}
Here, $A(\kappa,\tau)$ is given by \eqref{framematrix},
and $\mathbf T_g$ is determined by the target gate through
\eqref{eq:target-tangent}.

\subsection{Application of Pontryagin's Maximum Principle}

We will show in the sequel how Pontryagin's Maximum Principle (PMP) can be applied to determine necessary optimality conditions for the optimal control problem \eqref{eq:scqc-optimal-control}, resulting in a two-point boundary-value problem. The reader is referred to \cite{Boltyanskii1960,Boscain2021,Kirk2004,Troutman1996} for further details. 

First of all, fix $\mathbf B(0), \mathbf N(0), \mathbf B(T_g), \mathbf N(T_g)$ in \eqref{eq:scqc-optimal-control} so that $F(0), F(T_g) \in \mbox{SO}(3)$, and define the state vector $\mathbf x = \big(F, \kappa, \tau, \mathbf P, \mathbf Q \big) \in \mbox{SO}(3) \times \mathbb{R}^8$, $\mathbf f(\mathbf x, \mathbf u) = \big(A(\kappa, \tau)F, u, v, \mathbf T, \kappa \mathbf B \big)$ and $L(\mathbf{x},\mathbf{u}) = \frac{1}{2}(w_\kappa\kappa^2
+w_\tau\tau^2
+w_u u^2
+w_v v^2) \in \mathbb{R}$. Then, \eqref{eq:scqc-optimal-control} can be rewritten as   
\begin{equation}\label{ocp}
\begin{aligned}
\mathbf u^* &\in \operatorname*{arg\,min}_{\mathbf u} J = \int_0^{T_g} L(\mathbf{x},\mathbf{u})\,dt \\
\text{subject to:} \\
\quad
\dot{\mathbf x} &= \mathbf f(\mathbf x, \mathbf u ),\\
\mathbf x(0) &= \mathbf x_0, \\
\mathbf x(T_g) &= \mathbf x_g.
\end{aligned}
\end{equation}

The Hamiltonian associated to \eqref{ocp} is given by
\begin{equation}
\mathcal{H}\label{hamiltonian}
=
L(\mathbf{x},\mathbf{u})
+
\boldsymbol{\lambda}\cdot
\mathbf{f}(\mathbf{x},\mathbf{u}),
\end{equation}
where $\boldsymbol{\lambda}=(\boldsymbol\lambda_{\mathbf T}, \boldsymbol\lambda_{\mathbf N},\boldsymbol\lambda_{\mathbf B}, \lambda_\kappa, \lambda_\tau, \boldsymbol\lambda_{\mathbf P}, \boldsymbol\lambda_{\mathbf Q}) \in \mathbb{R}^{17}$ is the costate vector associated to the state vector $\mathbf x$, with $\boldsymbol\lambda_{\mathbf T}, \boldsymbol\lambda_{\mathbf B}, \boldsymbol\lambda_{\mathbf N}, \boldsymbol\lambda_{\mathbf P}, \boldsymbol\lambda_{\mathbf Q} \in \mathbb{R}^3$. The costate variables account for how variations in the state variables affect the functional $J$ along an optimal trajectory. As the frame $F = \left[ \mathbf T' \;\; \mathbf N' \;\; \mathbf B' \right]'$ is orthonormal, its admissible variations are infinitesimal rotations. A single rotation vector $\delta\boldsymbol\omega$ describes all three variations: $\delta\mathbf e=\delta\boldsymbol\omega\times\mathbf e$, for $\mathbf e\in\{\mathbf T,\mathbf N,\mathbf B\}$. Using the usual scalar triple product identity, the pairing of the frame costates with these variations becomes
\begin{equation}
\sum_{\mathbf e\in\{\mathbf T,\mathbf N,\mathbf B\}}
\boldsymbol\lambda_{\mathbf e}\cdot\delta\mathbf e
=\delta\boldsymbol\omega\cdot\mathbf m,
\end{equation}
where
\begin{equation}
\mathbf m=
\mathbf T\times\boldsymbol\lambda_{\mathbf T}
+\mathbf N\times\boldsymbol\lambda_{\mathbf N}
+\mathbf B\times\boldsymbol\lambda_{\mathbf B}.
\end{equation}
This construction is related to the rotational momentum map
used in geometric formulations in $\mathrm{SO}(3)$
\cite{Bloch2002}. Note that the 3 components of $\mathbf m \in \mathbb{R}^3$, expressed in fixed spatial coordinates, are sufficient to completely determine the costate pairing for admissible frame variations, since there are 6 redundant components in $(\boldsymbol\lambda_{\mathbf T}, \boldsymbol\lambda_{\mathbf N},\boldsymbol\lambda_{\mathbf B}) \in \mathbb{R}^9$. Indeed, three redundant directions are the components of each costate parallel to its own frame vector, and the other three redundant components   come from symmetric pairs of cross components. For example, adding $a\mathbf N$ to $\boldsymbol\lambda_{\mathbf T}$ and $a\mathbf T$ to $\boldsymbol\lambda_{\mathbf N}$ leaves $\mathbf m$ unchanged because the two cross products cancel; the same holds for the other two pairs. Therefore, we can replace the costate vector $\boldsymbol{\lambda}=(\boldsymbol\lambda_{\mathbf T}, \boldsymbol\lambda_{\mathbf N},\boldsymbol\lambda_{\mathbf B}, \lambda_\kappa, \lambda_\tau, \boldsymbol\lambda_{\mathbf P}, \boldsymbol\lambda_{\mathbf Q}) \in \mathbb{R}^{17}$ by the reduced costate vector $\boldsymbol{\lambda}_R=(\mathbf m, \lambda_\kappa, \lambda_\tau, \boldsymbol\lambda_{\mathbf P}, \boldsymbol\lambda_{\mathbf Q}) \in \mathbb{R}^{11}$, and \eqref{hamiltonian} yields the reduced Hamiltonian 
\begin{align}
\mathcal H_R{}=&\frac12(w_\kappa\kappa^2+w_\tau\tau^2+w_u u^2+w_v v^2)\nonumber\\
&+\kappa(\mathbf m+\boldsymbol\lambda_{\mathbf Q})\cdot\mathbf B
+\tau\mathbf m\cdot\mathbf T\nonumber\\
&+\boldsymbol\lambda_{\mathbf P}\cdot\mathbf T+\lambda_\kappa u+\lambda_\tau v,
\end{align}
where $w_\kappa$, $w_\tau \geq 0, w_u, w_v > 0$ are as in \eqref{functional}. The stationarity conditions $\displaystyle \partial\mathcal H_R/\partial u=\partial\mathcal H_R/\partial v=0$ provide the optimal controls \begin{equation}\label{optimalcontrols}
u^*=-\frac{\lambda_\kappa}{w_u},\qquad v^*=-\frac{\lambda_\tau}{w_v}.
\end{equation}
Let $\mathbf x^*(t)$ be the optimal state trajectory resulting from the optimal controls \eqref{optimalcontrols}. Then, the dynamics of $\boldsymbol{\lambda}_R(t)$ is determined by
\[
    \dot{\boldsymbol{\lambda}_R}(t) = \frac{\partial \mathcal H_R}{\partial \mathbf x}(\mathbf x^*(t), u^*(t), v^*(t), \boldsymbol{\lambda}_R(t)).    
\]
Therefore, the optimal state trajectory $\mathbf x^*(t) = \big(F^*(t),$ $\kappa^*(t),\tau^*(t), \mathbf P^*(t), \mathbf Q^*(t) \big) \in \mbox{SO}(3) \times \mathbb{R}^8$ and the reduced costate $\boldsymbol{\lambda}_R(t)=(\mathbf m(t), \lambda_\kappa(t), \lambda_\tau(t), \boldsymbol\lambda_{\mathbf P}(t), \boldsymbol\lambda_{\mathbf Q}(t)) \in \mathbb{R}^{11}$, $t \in [0, T_g]$, are governed by (recall \eqref{eq:scqc-optimal-control} and \eqref{ocp})
\begin{equation}
\begin{aligned}
\dot F^* &= A(\kappa^*,\tau^*)F^*,\\
\dot\kappa^* &= u^* = -\lambda_\kappa/w_u,\\
\dot\tau^* &= v^* = -\lambda_\tau/w_v,\\
\dot{\mathbf P}^* &= \mathbf T^*,\\
\dot{\mathbf Q}^* &= \kappa^*\mathbf B^*,\\[2pt]
\mathbf T^*(0) &= (0,0,1),\\
\mathbf T^*(T_g) &= \mathbf T_g \quad \mbox{as in \eqref{eq:target-tangent}},\\
F^*(0),\,F^*(T_g) &\in \mathrm{SO}(3)
\quad \text{(fixed)},\\
\kappa^*(0) &= \tau^*(0) = 0,\\
\kappa^*(T_g) &= \tau^*(T_g) = 0,\\
\mathbf P^*(0) &= \mathbf Q^*(0) = \mathbf 0,\\
\mathbf P^*(T_g) &= \mathbf Q^*(T_g) = \mathbf 0,\\[2pt]
\dot{\mathbf m}
&= -\mathbf T^*\times\boldsymbol\lambda_{\mathbf P}
   -\kappa^*\mathbf B^*\times\boldsymbol\lambda_{\mathbf Q},\\
\dot{\lambda}_\kappa
&= -w_\kappa\kappa^*
   -(\mathbf m+\boldsymbol\lambda_{\mathbf Q})
   \cdot\mathbf B^*,\\
\dot{\lambda}_\tau
&= -w_\tau\tau^*-\mathbf m\cdot\mathbf T^*,\\
\dot{\boldsymbol\lambda}_{\mathbf P}
&= \dot{\boldsymbol\lambda}_{\mathbf Q} = \mathbf 0.
\end{aligned}
\label{pmpequations}
\end{equation}
where $A(\kappa,\tau)$ is given by \eqref{framematrix} and $w_\kappa$, $w_\tau \geq 0, w_u, w_v > 0$ are as in \eqref{functional}, while the initial reduced costate vector $\boldsymbol{\lambda}_R(0)$ is unknown. The vector $\mathbf m(t)$ is expressed in fixed spatial coordinates and evolves as above, rather than rotated with the frame. These are the necessary optimality conditions determined by PMP, which is thus a two-point boundary-value problem associated to SCQC-based design. Hence, in order to solve the robust robust gate generation control problem here treated, it suffices to determine a numerical solution of the two-point boundary value problem \eqref{pmpequations}. 

It is worth noting that the choice of state and control variables adopted here is not unique. One possible alternative is to treat only $\kappa$ as a state variable, while using $u=\dot \kappa$ and $\tau$ as the two controls. In such a formulation, $\tau$ is determined directly by the stationarity condition of the PMP and therefore does not require separate endpoint conditions. Although this reduces the number of state variables, the two controls enter the optimization problem in an asymmetric manner, and we found this formulation to be less reliable in our numerical tests. We therefore retain both $\kappa$ and $\tau$ as state variables and use their derivatives as independent controls. With this choice, the endpoint values of $\tau$ must be treated consistently within PMP. Since $\tau$ is a state variable, it can either be fixed at the endpoints, as done here in \eqref{eq:scqc-optimal-control}, or left free with the corresponding transversality conditions imposed on its costate $\lambda_\tau$. The latter is a valid alternative formulation, but was also found to be less favorable in our numerical tests.

In the following section, we will solve the two-point boundary-value problem \eqref{pmpequations} using a shooting method, in which the initial reduced costate vector $\boldsymbol{\lambda}_R(0)$ is adjusted until the end-point constraints of $\mathbf x^*(t)$ are satisfied.

\section{\label{sec:shooting}Shooting Method}

We first describe how the two-point boundary-value problem \eqref{pmpequations} can be converted into a nonlinear root-finding problem and how the initial reduced costate vector $\boldsymbol{\lambda}_R(0)$ can be determined. We then discuss the construction of an initial solution and introduce a continuation strategy for the efficient synthesis of families of target gates.

\subsection{\label{subsec:shooting-formulation}Shooting Formulation}

The shooting method converts the two-point boundary-value
problem \eqref{pmpequations} into a sequence of initial-value
problems. We first specify the frame boundary data
used in the numerical implementation.

As discussed in Sec.~\ref{sec:scqc}, the target gate fixes
the terminal tangent through \eqref{eq:target-tangent},
while the remaining $Z$ rotation can be compensated
virtually. Therefore, the target-gate condition does not
uniquely determine the terminal normal and binormal.
We exploit this freedom when selecting the fixed frame
boundary data required by the shooting formulation.

For a prescribed unit tangent $\mathbf T_{\rm ref}$,
orthonormality requires the normal to be a unit vector in
the plane perpendicular to $\mathbf T_{\rm ref}$. Given a
positively oriented reference orthonormal frame
$(\mathbf T_{\rm ref},\mathbf N_{\rm ref},\mathbf B_{\rm ref})$,
the vectors $\mathbf N_{\rm ref}$ and $\mathbf B_{\rm ref}$
form an orthonormal basis of this plane. Hence, every
admissible normal can be written as
$\mathbf N_\alpha=a\mathbf N_{\rm ref}+b\mathbf B_{\rm ref}$,
with $a^2+b^2=1$. Setting $a=\cos\alpha$ and $b=-\sin\alpha$,
the binormal is uniquely determined by
$\mathbf B_\alpha=\mathbf T_{\rm ref}\times\mathbf N_\alpha$.
Thus,
\begin{align}
\mathbf T_\alpha&=\mathbf T_{\rm ref},\nonumber\\
\mathbf N_\alpha&=\cos\alpha\,\mathbf N_{\rm ref}
                 -\sin\alpha\,\mathbf B_{\rm ref},\nonumber\\
\mathbf B_\alpha&=\sin\alpha\,\mathbf N_{\rm ref}
                 +\cos\alpha\,\mathbf B_{\rm ref}.
\label{eq:frame-angle}
\end{align}
This parametrization preserves orthonormality and positive
orientation and describes all frames with the prescribed
tangent, with $\alpha$ defined modulo $2\pi$. It can be
applied at either endpoint, with angles denoted by
$\alpha_0$ and $\alpha_{T_g}$.

One possible formulation allows the initial normal and binormal to vary through $\alpha_0$, while leaving their terminal orientation unspecified. Rodrigues' updates preserve orthonormality throughout integration, and the remaining $z$ rotation can be compensated virtually. If the terminal orientation is treated as free in the optimal-control problem, PMP requires an additional transversality condition. Alternatively, introducing $\alpha_{T_g}$ explicitly defines fixed terminal frame data for each selected angle, but varying both endpoint angles adds two angular parameters to the shooting problem and requires accounting for the initial pulse phase.

In our numerical comparisons, retaining only the terminal angle provided the most favorable balance between reference-solution preparation, continuation efficiency, and implementation simplicity. We fix $\Phi(0)=0$ as the reference phase of the control, aligning the initial control direction with the $x$ axis. With $U_0(0)=I$, this phase convention defines the initial reference frame
\begin{equation}
\begin{aligned}
\mathbf T(0)&=(0,0,1),\\
\mathbf N(0)&=(0,1,0),\\
\mathbf B(0)&=(-1,0,0).
\end{aligned}
\end{equation}
We then set $\alpha_0=0$, so that no additional rotation is applied to this frame. 

To construct the terminal target for the shooting method,
we first define a nominal reference frame through the
adjoint representation of the target operation:
\begin{equation}
\begin{pmatrix}
-\mathbf B_{\rm ref}'\\
\mathbf N_{\rm ref}'\\
\mathbf T_{\rm target}'
\end{pmatrix}
=R_{U_{\rm target}}\mathcal R_F(0).
\label{eq:nominal-target-frame}
\end{equation}
The tangent $\mathbf T_{\rm target}$ is the target tangent
defined in \eqref{eq:target-tangent}.

We then apply \eqref{eq:frame-angle} to the reference
frame defined in \eqref{eq:nominal-target-frame}, with
$\alpha=\alpha_{T_g}$. We denote the resulting normal and
binormal by $\mathbf N_{\rm target}(\alpha)$ and
$\mathbf B_{\rm target}(\alpha)$, respectively.

The angle $\alpha$ is treated as a parameter of the terminal boundary data, rather than as a free endpoint variable in the variational problem. During the reference solve, it is adjusted together with the initial costates to satisfy the shooting constraints. During subsequent corrections, it is held at its predicted value. No transversality condition associated with $\alpha$ is imposed in either stage, since the procedure selects terminal boundary data rather than minimizing the trajectory cost over the terminal orientation.

The terminal angular freedom changes the residual $z$ rotation but preserves the desired gate after virtual phase correction. With $\Delta=0$, the corrected operation is (see \eqref{correctionrotation}--\eqref{thetarotation})
\begin{equation}
U_{\rm corrected}=R_z[-\alpha-\Phi(T_g)]U_{\rm target},
\label{eq:alpha-correction}
\end{equation}
up to a global phase.

The shooting parameters are
\begin{equation}
\mathbf z=(\mathbf m(0),\lambda_\kappa(0),\lambda_\tau(0),
\boldsymbol\lambda_{\mathbf Q},\boldsymbol\lambda_{\mathbf P},\alpha)
\in\mathbb R^{12}.
\label{eq:shooting-vector}
\end{equation}
For a given trial vector, the complete state--costate system is initialized at $t=0$ and propagated forward to $t=T_g$. The resulting terminal state is compared with the prescribed boundary conditions, defining
\begin{equation}
\mathbf F(\mathbf z,\theta)=
\begin{pmatrix}
F(T_g)-F_g(\alpha)\\
\kappa(T_g)\\
\tau(T_g)\\
\mathbf Q(T_g)\\
\mathbf P(T_g)
\end{pmatrix}
\in\mathbb R^{17}.
\end{equation}

Here $F_g(\alpha)$ is the prescribed terminal frame with rows $\mathbf T_{\rm target}'$, $\mathbf N_{\rm target}(\alpha)'$, and $\mathbf B_{\rm target}(\alpha)'$. The first block contributes the nine entries of the terminal-frame difference to the residual vector.

Equivalently, the terminal-frame condition can be imposed through the relative rotation $F_g(\alpha)^{\mathsf T}F(T_g)=I_3$. The corresponding matrix residual has the same Frobenius norm as the frame difference above. We retained the frame difference because it performed slightly better in our numerical comparison.

The shooting solver seeks $\mathbf F(\mathbf z,\theta)=\mathbf0$. For the reference solution, the initial guess is not informed by a nearby accepted trajectory. We therefore use a bounded trust-region reflective (TRF) algorithm, which enforces the prescribed parameter limits throughout the search. All twelve shooting parameters are adjusted using a two-point finite-difference Jacobian.

Subsequent continuation and interpolation queries instead provide informed initial estimates, so the remaining task is a local correction. We use Levenberg--Marquardt for this small, dense nonlinear least-squares problem, adjusting eleven components while holding $\alpha$ at its predicted value. This choice exploits the computational efficiency of LM for unconstrained corrections. Since LM does not enforce bounds internally, both the parameter limits and residual tolerances are checked before accepting a solution. For each trial shooting vector, we integrate the state--costate system in \eqref{pmpequations}, in which the stationary controls $u^*=-\lambda_\kappa/w_u$ and $v^*=-\lambda_\tau/w_v$ are already substituted into the evolution equations for $\kappa^*$ and $\tau^*$, respectively.

A key feature of our implementation is the treatment of the Frenet--Serret frame during the forward propagation. Rather than numerically integrating its differential equations directly, we exploit the fact that the matrix $A(\kappa,\tau)$ is skew-symmetric. Freezing $\kappa$ and $\tau$ over an integration interval $\Delta t$, the frame evolves according to a finite rotation generated by this matrix,
\begin{equation}
F(t+\Delta t)
=
\exp\!\left[A(\kappa,\tau)\Delta t\right]
F(t),
\end{equation}
For piecewise-constant $\kappa$ and $\tau$ over each integration interval, the matrix exponential can be evaluated analytically using Rodrigues' rotation formula \cite[Section 3.2.3]{Lynch2017}. Defining
\begin{equation}
\boldsymbol{\omega}
=
\begin{pmatrix}
-\tau\\
0\\
-\kappa
\end{pmatrix},
\end{equation}
the matrix $A$ can be interpreted as the cross-product matrix associated with $\boldsymbol{\omega}$. Its magnitude is
\begin{equation}
\|\boldsymbol{\omega}\|
=
\sqrt{\kappa^2+\tau^2}.
\end{equation}
The corresponding finite rotation over a step $\Delta t$ is then given by Rodrigues' formula,
\begin{equation}
\exp\!\left[A\Delta t\right]
=
I
+
\frac{\sin(\|\boldsymbol{\omega}\|\Delta t)}
{\|\boldsymbol{\omega}\|}A
+
\frac{1-\cos(\|\boldsymbol{\omega}\|\Delta t)}
{\|\boldsymbol{\omega}\|^2}A^2.
\end{equation}

Thus, the Frenet--Serret frame is propagated through exact frozen-coefficient rotations within each step. The scalar states and costates use explicit Euler updates; $\mathbf m$ uses the updated frame and the old curvature, while $\mathbf P$ and $\mathbf Q$ use trapezoidal sums. The complete scheme remains first order and subject to discretization error, despite preserving frame orthonormality. This procedure is repeated for each trial set of initial costates during the shooting iterations.

From a computational perspective, each residual evaluation consists of a forward integration of the reduced state--costate system. The reference search adjusts twelve parameters, while a fixed-$\alpha$ correction adjusts eleven. Obtaining suitable initial guesses is therefore an important part of the numerical procedure. We address this issue in the following subsection.

\subsection{\label{subsec:initial-guess}Initial Guess Construction}

In practice, convergence of the shooting procedure is sensitive to the initial costates. We first solve the complete shooting problem for a representative target gate, starting from a pseudorandom guess. The bounded TRF solve includes $\alpha$ among its twelve adjustable parameters.

The reference solution obtained above provides a starting point for constructing initial guesses for a broader set of target operations. For a single qubit, an arbitrary unitary can be decomposed, up to a global phase, as
\begin{equation}
U_3(\theta,\phi,\lambda)
=
R_z(\lambda)R_x(\theta)R_z(\phi).
\end{equation}
The $R_z$ rotations can be implemented virtually through frame tracking and therefore do not require a physical control pulse. Consequently, the physical pulse synthesis can be restricted, without loss of generality, to rotations $R_x(\theta)$ about a fixed axis. This reduces the construction of initial guesses for arbitrary single-qubit gates to the determination of suitable shooting parameters as a function of the single rotation angle $\theta$.

We follow a continuous sequence of solutions from this reference using tangent prediction and adaptive correction \cite{BaderKunkel1989}. Differentiating $\mathbf F(\mathbf z(\theta),\theta)=0$ at an accepted solution gives
\begin{equation}
J_i\mathbf t_i=-\mathbf F_{\theta,i},\qquad
J_i=\partial_{\mathbf z}\mathbf F(\mathbf z_i,\theta_i),
\end{equation}
and the next initial guess is
\begin{equation}
\mathbf z_{\rm pred}=\mathbf z_i+\Delta\theta\,\mathbf t_i.
\end{equation}
The Jacobian is computed by complex-step differentiation of the discrete residual \cite{Martins2003},
\begin{equation}
(J_i)_{:j}\simeq\frac{\operatorname{Im}\mathbf F(\mathbf z_i+i\eta\mathbf e_j,\theta_i)}{\eta},
\qquad\eta=10^{-25}.
\end{equation}
The integration kernels preserve complex arithmetic. The angle derivative uses a centered difference with step $10^{-6}$. To account for parameter scales and redundant residual entries, we solve the tangent equation in scaled minimum-norm least squares: with $D_i=\operatorname{diag}\{\max(|z_{ij}|,1)\}$, solve $J_iD_i\mathbf y_i=-\mathbf F_{\theta,i}$ using a relative singular-value cutoff of $10^{-12}$, then set $\mathbf t_i=D_i\mathbf y_i$.

Two independent sweeps start from the reference, toward the lower and upper endpoints. Each prediction is corrected by Levenberg--Marquardt with a complex-step Jacobian, holding $\alpha$ at its predicted value. A rejected step is halved and retried from the last accepted solution. After acceptance, the step grows by $20\%$ for fewer than ten corrector function evaluations, decreases by $30\%$ for more than twenty-five, and is otherwise retained. Requested intermediate targets and prescribed step limits are respected. Near zero, the angular step is also capped relative to the current angle, allowing a denser set of solutions where the parameters vary rapidly.

Only accepted solutions are retained. Each component of the twelve-component shooting vector is then interpolated using a shape-preserving piecewise cubic Hermite interpolating polynomial (PCHIP) \cite{FritschButland1984}. For a new target angle $\theta^*$ inside the covered interval, the interpolated vector supplies the initial estimate. Its first eleven components are corrected while $\alpha$ remains fixed at its interpolated value. The complete terminal and robustness conditions and the parameter bounds are verified before acceptance.

The resulting procedure provides an efficient initialization scheme for new rotation angles. Reference and continuation preparation are performed once, after which the fixed interpolator is reused for subsequent queries. Interpolation predicts initial guesses rather than replacing the shooting correction. No extrapolation outside the covered interval is assumed. The performance of this procedure is evaluated in the following section.

\begin{figure*}[t]
\centering
\includegraphics[width=1\linewidth]{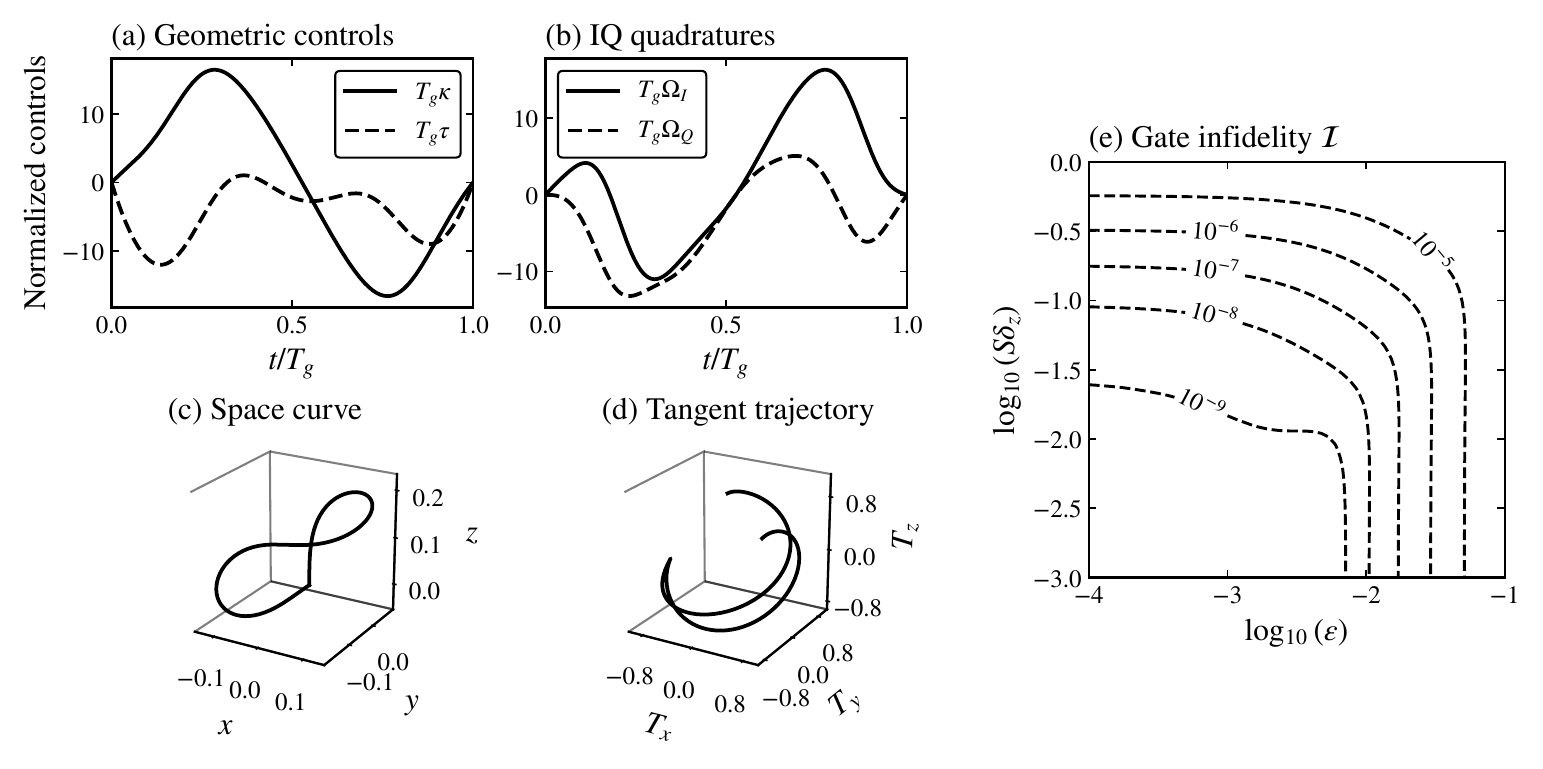}
\caption{Control, geometric, and propagation representations of the robust $R_x(\pi/2)$ reference pulse. (a) Curvature and torsion; (b) IQ quadratures, reconstructed from $\Omega=\kappa$ and $\Phi(t)=\int_0^t\tau(t')\,dt'$; (c) space curve; (d) tangent trajectory on the unit sphere; (e) average gate infidelity under simultaneous quasistatic amplitude and detuning errors, with the nominal virtual correction fixed. The noise map uses a $60\times60$ logarithmic grid covering $\epsilon\in[10^{-4},10^{-1}]$ and $T_g\delta_z\in[10^{-3},1]$.}
\label{fig:curve}
\end{figure*}

\section{\label{sec:validation}Numerical Validation}

We evaluate the proposed shooting-based synthesis method with respect to three aspects: the construction of a converged reference solution, the behavior of the shooting parameters as a function of the target rotation angle, and the computational cost and accuracy of the resulting synthesis. All experiments were performed on a laptop equipped with an Intel Core i7-13620H processor and 32~GB of RAM. The experiments were run under Windows. The implementation uses Python 3.14.7, with SciPy 1.18.1 and Numba 0.67.0.

\subsection{Reference Solution}

The first step of the proposed procedure is to obtain a converged reference solution from which reliable initial guesses can be constructed for other rotation angles. We selected the representative target rotation $R_x(\pi/2)$ and initialized the shooting parameters randomly. The normalized duration was $T_g=1$, with 1,000 integration intervals. In the convention of the objective functional presented above, the weights were
\begin{equation}
w_\kappa=w_\tau=2,\qquad w_u=w_v=0.2.
\end{equation}

In the recorded reference-preparation run, the complete twelve-parameter problem was solved by bounded TRF with a two-point Jacobian in $6.956$~s, with total residual norm $6.047\times10^{-13}$. All terminal and robustness conditions were retained from the start. The initial reference search is accounted for separately from continuation and interpolation queries.

The first eleven parameters were restricted to $[-10^4,10^4]$, and $\alpha$ to $[0,2\pi]$. Acceptance required finite parameters and residuals, satisfied bounds, and values below $10^{-6}$ for each of the following: the norm of the nine frame residuals, $|\kappa(T_g)|$, $|\tau(T_g)|$, $\|\mathbf Q(T_g)\|$, and $\|\mathbf P(T_g)\|$.

Figure~\ref{fig:curve} shows the reference solution for $R_x(\pi/2)$, including the curvature and torsion, physical IQ quadratures, spatial curve $\mathbf r(t)$, and tantrix $\mathbf T(t)$. These quantities illustrate the relation between the geometric representation and the physical control used to implement the gate. Reconstruction of the curves uses the same geometric integration convention as shooting and provides a consistency check.

In addition to the terminal and robustness residuals, we evaluated the nominal gate fidelity of the reference solution by propagating its reconstructed physical IQ controls. On each interval, the IQ quadratures were averaged over its endpoints and the resulting constant two-level Hamiltonian was exponentiated. The nominal virtual correction in \eqref{eq:alpha-correction} was then applied. We evaluate the corrected operation using \eqref{eq:gate-infidelity}.

For the same reference pulse, Fig.~\ref{fig:curve}(e) evaluates simultaneous quasistatic multiplicative amplitude error and additive detuning. The propagation uses
\begin{equation}
H_{\epsilon,\delta_z}=\frac12\left[(1+\epsilon)(\Omega_I\sigma_x+\Omega_Q\sigma_y)+\delta_z\sigma_z\right].
\end{equation}
The nominal virtual phase correction remains fixed throughout the scan, without noise-dependent adjustment or reoptimization. This finite-error propagation complements the first-order geometric robustness conditions; the complete noise map is evaluated for the reference pulse.

\subsection{Angular Continuation and Interpolation}

Once the reference solution was obtained, we investigated the dependence of the shooting parameters on the target rotation angle. The requested grid consisted of 100 equally spaced angles in $[0,\pi]$, with the exact reference angle added. Two independent tangent-continuation sweeps started from the corrected reference. Adaptive steps inserted intermediate solutions, yielding 1,175 accepted points including the reference. Three rejected attempts were recovered by reducing the step. Both endpoints, including $\theta=0$, were reached. Continuation additionally required a total residual norm below $10^{-8}$.

The initial angular step was $0.05$~rad, with maximum $0.1$~rad, minimum $10^{-8}$~rad, and at most twenty retries. For $10^{-5}<|\theta|<0.05$, the step was capped at $0.015|\theta|$; below $10^{-5}$ this cap was released to attempt the endpoint. The fixed-$\alpha$ corrector used Jacobian-based scaling, termination tolerances $10^{-12}$, and at most sixty function evaluations.

\begin{figure*}[t]
\centering
\includegraphics[width=1\linewidth]{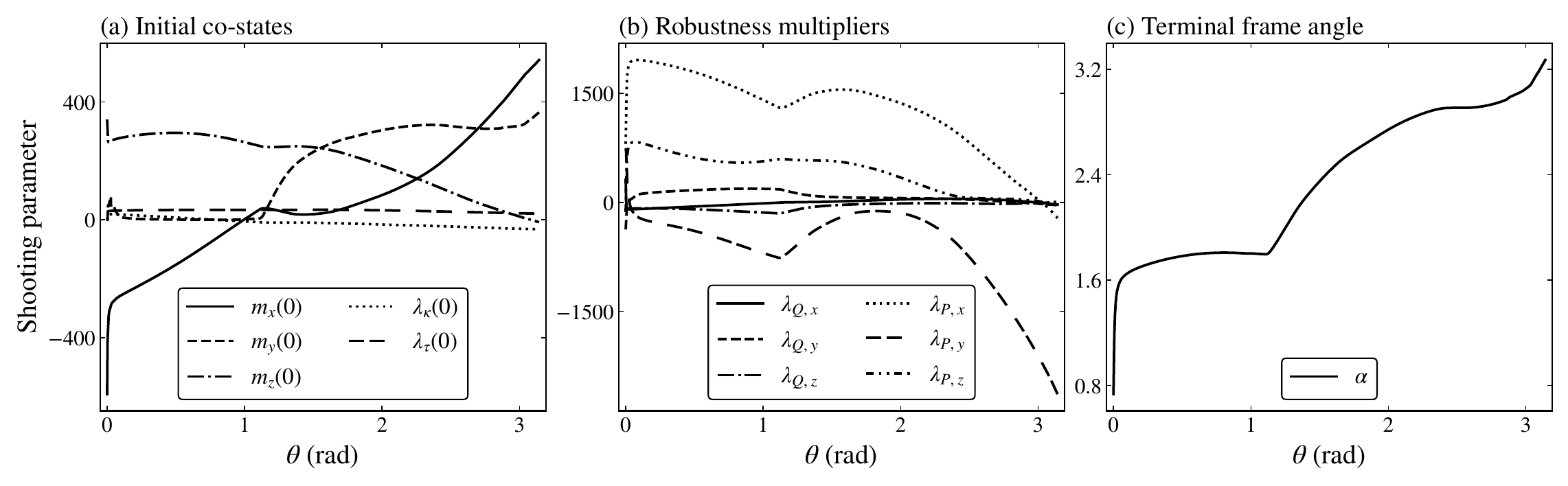}
\caption{Interpolated shooting parameters as functions of the target rotation angle $\theta$. (a) $\mathbf m(0)$, $\lambda_\kappa(0)$, and $\lambda_\tau(0)$; (b) $\boldsymbol\lambda_{\mathbf Q}$ and $\boldsymbol\lambda_{\mathbf P}$; (c) $\alpha$. PCHIP interpolation is constructed from 1,175 accepted continuation points covering $[0,\pi]$.}
\label{fig:interpolation}
\end{figure*}

Figure~\ref{fig:interpolation} shows the interpolated shooting parameters along the continuation sequence. The parameters vary rapidly near $\theta=0$, making this region particularly challenging for continuation initialized directly from the preceding solution. Tangent prediction combined with adaptive step reduction enabled the procedure to follow this rapid variation and reach the zero-angle endpoint.

The total continuation time was $88.049$~s, including correction of the base, tangent prediction, correction attempts, and sweep control. For the 1,174 accepted corrections after the base whose optimizer times were recorded, the mean was $54.488$~ms, the median $47.024$~ms, and the 95th percentile $84.341$~ms. These times describe preparation of the reusable interpolation base.

\subsection{Validation of Interpolation-Initialized Queries}

To evaluate the interpolated initialization, we generated 2,000 equally spaced angles over $[0,\pi]$. Each initial shooting vector was obtained directly from the same fixed PCHIP interpolator, rather than from a previously solved neighboring angle. An eleven-parameter Levenberg--Marquardt correction held $\alpha$ at its interpolated value. All 2,000 queries satisfied the acceptance criteria. The grid includes endpoints and may coincide with other continuation points; it is therefore not an entirely held-out set.

\begin{figure}[h]
\centering
\includegraphics[width=\linewidth]{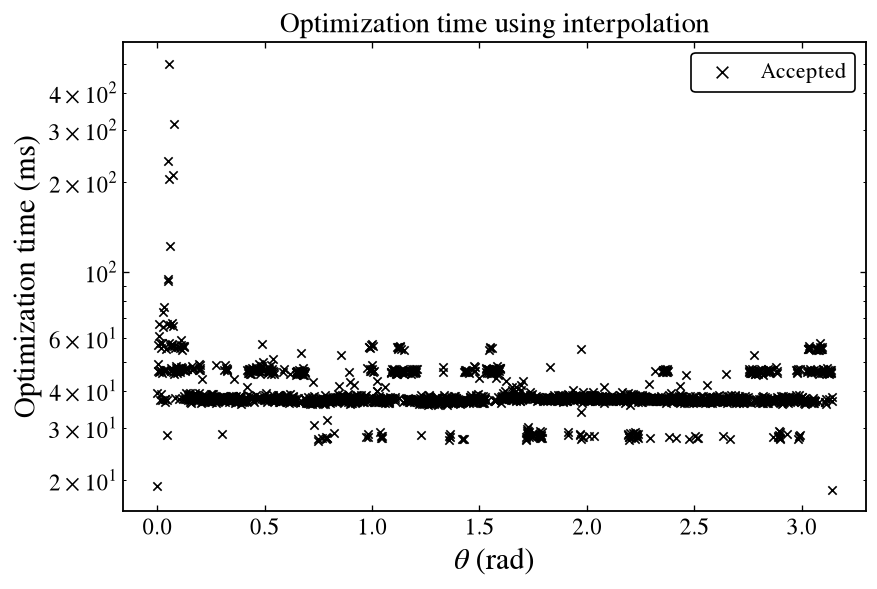}
\caption{Optimizer time for each of the 2,000 interpolation-initialized queries on $[0,\pi]$. Cross markers indicate individual queries, with a logarithmic time axis. All queries were accepted. Times measure the optimizer call only and exclude reference and library preparation.}
\label{fig:query-times}
\end{figure}

Figure~\ref{fig:query-times} shows the optimizer time for each angle. The mean was $39.945$~ms, the median $37.575$~ms, and the 95th percentile $48.043$~ms. The minimum was $18.571$~ms at $\theta=\pi$, and the maximum was $501.135$~ms at $\theta=0.05343379$~rad. The sum of the optimizer times was $79.891$~s. Typical corrections therefore took tens of milliseconds, with the observed half-second peak retained in the reported results.

Individual times measure only the call to \texttt{least\_squares}. They exclude interpolation, initial residual and Jacobian evaluations before timing, control reconstruction, and quantum propagation. Compilation is performed before the individual optimizer timer. The printed times have resolution $0.001$~ms. These measurements describe one recorded execution, rather than repeated-run uncertainty or complete pulse-generation latency.

Using the same quantum-propagation procedure and \eqref{eq:gate-infidelity}, we evaluated all 2,000 accepted query pulses. The mean infidelity was $5.526\times10^{-10}$, the median $7.049\times10^{-10}$, and the maximum $9.313\times10^{-10}$ at $\theta=2.22221711$~rad. All tested queries therefore had nominal infidelity below $10^{-9}$ on the stated discretization. This propagation check is separate from evaluation of the shooting residual; it does not constitute a continuous-time integration-error bound.

The resulting performance demonstrates the main computational advantage of the proposed approach. The computationally expensive continuation stage is performed only once, after which the interpolated shooting parameters provide informed initial conditions for new target rotations. Consequently, the cost of a query is reduced to the local correction of an interpolated initial guess rather than a new search from random costates.

\section{\label{sec:conclusion}Conclusion}

In this work, we addressed the problem of reducing the computational cost of robust quantum gate synthesis. Motivated by applications requiring robust controls to be generated at compile time \cite{Lussi2025}, we proposed a semi-analytical approach based on the SCQC framework and PMP. By combining a two-point boundary value problem-based shooting method, Rodrigues' formula, and angular continuation with interpolation, we obtained robust controls without relying on costly high-dimensional pulse optimization. The method successfully corrected 2,000 single-qubit rotation queries using a fixed PCHIP library, with a mean optimizer time of $39.945$~ms and a maximum nominal infidelity of $9.313\times10^{-10}$ after ideal virtual phase correction on the stated discretization. Reference and continuation preparation were measured separately.

Compared with approaches such as BARQ, the proposed method does not enforce gate fixing and dephasing robustness exactly by construction; instead, these conditions are satisfied numerically by means of the referred two-point boundary value problem. Nevertheless, quantum propagation confirmed the nominal accuracy of the synthesized gates after virtual phase correction, while finite-error simulations demonstrated the robustness of the reference pulse against quasistatic amplitude and dephasing errors. Furthermore, the smooth and structured behavior observed in the controls and shooting parameters suggests the existence of an underlying analytical structure, which we leave for future investigation.

Our main future direction is to incorporate this synthesis procedure into dynamic quantum compilation \cite{Lussi2025}. The low generation cost makes it possible to generate controls at compile time, allowing each gate instance to be optimized in duration and synthesized according to its specific circuit context. Thus, identical logical gates at different positions in a circuit could be implemented by different physical controls, moving robust pulse synthesis from an offline optimization step toward a dynamic, context-aware compilation process. We also plan to extend the synthesis method to two-qubit gates and incorporate the suppression of leakage out of the computational subspace into the control formulation.

\begin{acknowledgments}
This work was partially supported by Fundação de Amparo à Pesquisa e Inovação do Estado de Santa Catarina (FAPESC) under Public Call No. 24/2025 and the National Institute of Science and Technology for Applied Quantum Computing (INCT-CQA) under Grant No. 408884/2024-0.
\end{acknowledgments}

\bibliography{apssamp}

\end{document}